\documentclass[conference]{IEEEtran}
\IEEEoverridecommandlockouts
\usepackage{cite}
\usepackage{amsmath,amssymb,amsfonts}
\usepackage{algorithmic}
\usepackage{adjustbox}
\usepackage{graphicx}
\usepackage[para,online,flushleft]{threeparttable}
\usepackage{textcomp}
\usepackage[dvipsnames]{xcolor}
\def\BibTeX{{\rm B\kern-.05em{\sc i\kern-.025em b}\kern-.08em
    T\kern-.1667em\lower.7ex\hbox{E}\kern-.125emX}}
\begin{document}

\bstctlcite{BSTcontrol}

\title{Revealing Untapped DSP Optimization Potentials for FPGA-Based Systolic Matrix Engines}

\author{Jindong Li$^{1, 2, 4}$ \ \ Tenglong Li$^{1, 2, 4}$ \ \ Guobin Shen$^{1, 2, 3}$ \ \ Dongcheng Zhao$^{1,2}$ \ \  Qian Zhang$^{1, 2, 4}$ \ \ Yi Zeng$^{1, 2, 3, 4, 5}$\\
$^1$Brain-inspired Cognitive Intelligence Lab, Institute of Automation, Chinese Academy of Sciences\\ 
$^2$ Center for Long-term Artificial Intelligence \\
$^3$ School of Future Technology, $^4$ School of Artificial Intelligence, University of Chinese Academy of Sciences  \\ 
$^5$ Key Laboratory of Brain Cognition and Brain-inspired Intelligence Technology, Chinese Academy of Sciences\\ 
{\{lijindong2022, litenglong2023, shenguobin2021,}{zhaodongcheng2016, q.zhang, yi.zeng\}@ia.ac.cn}
\thanks{Corresponding author: q.zhang@ia.ac.cn and yi.zeng@ia.ac.cn.}
}

\maketitle

\begin{abstract}
Systolic architectures are widely embraced by neural network accelerators for their superior performance in highly parallelized computation. The DSP48E2s serve as dedicated arithmetic blocks in Xilinx Ultrascale series FPGAs and constitute a fundamental component in FPGA-based systolic matrix engines. Harnessing the full potential of DSP48E2s in architectural design can result in significant performance enhancements for systolic architectures on Ultrascale series FPGAs.
This paper unveils several previously untapped DSP optimization techniques capable of further enhancing FPGA-based systolic matrix engines.
We apply these techniques to two well-known systolic architectures: Google TPUv1 and Xilinx Vitis AI DPU.
With the proposed techniques, our design achieves substantial resource and power reduction compared to the open-source TPUv1 FPGA implementation and the Vitis AI DPU implementation in the same parallelism setting.
We also demonstrate the applicability of our techniques to neuromorphic hardware for supporting spiking neural network acceleration.
\end{abstract}


\section{Introduction}
Recent years have witnessed the development of domain-specific hardware architectures tailored for deep learning.  Systolic architectures have demonstrated their effectiveness and efficiency in handling matrix multiplication during neural network inference.

In FPGA-oriented design, building a high-performance systolic matrix engine is a non-trivial endeavor. While a naive or high-level synthesis (HLS) implementation may function, it falls short of unlocking the maximum performance potential. Achieving the pinnacle of performance for a systolic matrix engine on an FPGA demands an in-depth understanding of the FPGA's resource characteristics. In the case of the UltraScale series FPGA, harnessing the full potential of the basic systolic array's building block, DSP48E2\cite{xilinx2021ultrascale}, becomes as a paramount consideration.
A significant amount of research has concentrated on DSP-centric optimization to unlock the performance improvement potential\cite{fu20178}\cite{han2020convolutional}\cite{sommer2022dsp}\cite{zhang2023uint}\cite{li2023fireflyv1}. Several techniques have already become common practices in contemporary neural network accelerator designs.

Keeping pace with the evolution of neural network algorithms, new domain-specific hardware architectures have been proposed rapidly. Yet, the systolic architecture continues to play a crucial role as the foundational computing backend.
We recognize the necessity to revisit the DSP48E2 functionalities to determine if further optimization of FPGA-based systolic engines is possible. Our findings in this paper confirm that there are indeed several untapped DSP optimization potentials for FPGA-based systolic engines that can be beneficial to the FPGA-based accelerator research society.

In this paper, we focus on two classic systolic-based neural network accelerators: the Google TPUv1\cite{jouppi2017datacenter} and Xilinx Vitis AI DPU\cite{xilinx2021dpu}. We present pratical DSP48E2 techniques that can enhance the performance of the systolic matrix engine in these two accelerators. Our contributions are listed as follows:

1) We propose an in-DSP operand prefetching techniques applicable to the weight stationary (WS) systolic engine. With such technique, our implementation shows substantial resources reduction and frequency improvement compared to currently widely adopted open-source TPUv1-like design.

2) We delve deeply into the systolic array implementation of the commercially encrypted Vitis AI DPU, identifying existing drawbacks in its design. Subsequently, we propose an in-DSP multiplexing technique and design a ring accumulator that can further reduce resource consumption and power consumption compared to the original official design.

3) We also demonstrate the application of our methods to neuromorphic hardware by offering enhanced implementations of the systolic-based spiking neural network (SNN) accelerators, FireFly\cite{li2023fireflyv1}.

\section{Related Works}

DSP-based techniques play an important role in FPGA-based accelerators since DSP blocks serve as the fundamental computing hard blocks, yet being limited and scarce. Our focus in this paper is specifically on the UltraScale series FPGAs, which are widely favored in deep learning applications.

A common practice in neural network accelerator designs involves utilizing the wide input arithmetic unit within the DSP48E2 in UltraScale series FPGAs to enable SIMD low bit-width operations in quantized neural networks\cite{chen2021hardware}\cite{lee2018double}.
Xilinx introduced a method that pack two 8-bit integer multiplications sharing a same operand into a single DSP48E2\cite{fu20178}. They further proposed another approach that pack the cross products between two pair of 4-bit operands into a single DSP48E2\cite{han2020convolutional}. Zhang et al. proposed UInt-DSP6 packing and packed two more 4-bit multiplications into the DSP48E2 with appropriate overlaping in convolution application\cite{zhang2023uint}. Sommer et al. generalized the multiplication packing technique and introduced an overpacking strategy\cite{sommer2022dsp}. FireFly\cite{li2023fireflyv1} utilizes the SIMD mode of DSP48E2 to perform multiple synaptic operations for neuromorphic acceleration.

Another common practice in neural network accelerator designs involves leveraging the cascading paths in DSP48E2 for constructing systolic arrays. The DSP48E2 comprises three cascade paths: two to the input A and B ports and one to the output P port.
Xilinx's white paper suggests utilizing the partial sum cascaded chain at the output P port of the DSP48E2 for performing 8-bit integer dot product operations\cite{fu20178}.
Additionally, the Xilinx DSP48E2 datasheet recommends the adder-chain implementation over the adder-tree approach for FIR filter applications requiring high speed\cite{xilinx2021ultrascale}.
Samajdar et al. optimally utilize the cascade chain at the input A and B ports in the DSP48E2 to support fixed-mode convolution\cite{samajdar2019scaling}.

The DSP48E2's superior Fmax performance is also commonly utilized in neural network accelerator designs. The peak clock rate of DSP48E2 can exceed 700MHz for a device of the fastest speed grade\cite{sheet2017virtex}. To fully utilize its superior Fmax performance, the Xilinx DSP48E2 datasheet recommends employing the DSP48E2 in a time-multiplexed manner\cite{xilinx2021ultrascale}. Following this guideline, Vitis AI DPUCZDX8G adopts a DSP double data rate (DDR) technique to enhance its systolic engine performance. FireFly v2\cite{li2024fireflyv2} adopts the DDR technique on a neuromorphic SNN accelerator. Additionally, Wu et al. proposed a DSP supertile concept by utilizing the near DSP distributed RAM for fast operand fetching\cite{wu2017high}.

While the potential for DSP48E2 optimization opportunities appears to have reached its end, this paper unveils several untapped tricks that can still greatly enhance performance.

\section{DSP48E2 Overview}

\begin{figure}
    \centering
    \includegraphics[width=1.0\linewidth]{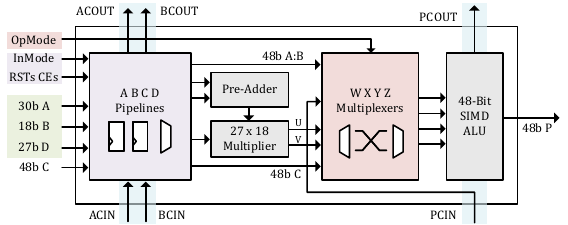}
    \caption{DSP48E2 Overview. \color{JungleGreen}{Green: four wide input ports.} \color{Purple}{Purple: two flexible input pipelines.} \color{Maroon}{Orange: four dynamic multiplexers.} \color{RoyalBlue}{Blue: three cascade paths.}}
    \label{fig:dsp48e2}
    \end{figure}

DSP48E2 serve as fundamental computing components in FPGAs, providing extensive parallelism opportunities for handling intensive computing workloads in neural network acceleration in Xilinx Ultrascale series FPGAs.

The DSP48E2 primarily comprises a 27-bit pre-adder, a $27\times 18$-bit multiplier, and an SIMD 48-bit accumulator, shown in Fig.\ref{fig:dsp48e2}. However, some circuits in the DSP48E2 do not directly contribute to arithmetic computations; instead, they play a crucial role as essential components for reconfigurable functionalities.
The \textit{flexible input pipelines} inside the DSP48E2 feature two distinct pipeline registers with individual clock enables, along with a dynamic selector present on both the A and B input ports, enabling various input configurations.
The four \textit{wide-bus multiplexers} route data from different ports to the four-input 48-bit ALU, enabling various dynamic user-controlled operating modes.
The \textit{dedicated cascade paths} enable direct connections between adjacent DSP48E2s in the same column, enabling high-speed systolic-based applications.

In this paper, we uncover several DSP48E2 techniques by exploring these often-overlooked components in DSP48E2.

\section{Enhancing Systolic Engine of TPUv1 on FPGA}

Google TPUv1\cite{jouppi2017datacenter} is a Tensor Processing Unit (TPU) designed for neural network machine learning. It incorporates a $256\times 256$ 8-bit WS Multiply-Accumulate (MAC) matrix unit, delivering a peak throughput of 92 TOP/s.
In the classic WS systolic array design like TPUv1, weight data are fetched and cached near the PEs in advance, keeping stationary until the arrival of new sets of weight data. Input data flows horizontally into the systolic array staging into the next PE, while the partial sums flow vertically out of the array, accumulating along the way.
This architecture proves to be efficient for matrix multiplication, which is the computing backend for the widely used nn.Linear layer and nn.Conv2d layer.

tinyTPU\cite{tinytpugithub} is the most widely adopted open-source FPGA-based TPUv1 design\cite{arora2021tensor}\cite{yang2022bp}\cite{he2020sparse}. It large follows TPUv1 design but using a smaller systolic matrix size of five configurable choices ranges from $6\times 6$ to $14\times 14$ that can fit into edge FPGAs. 
Libano's systolic array generator represents a state-of-the-art FPGA-based TPUv1-like systolic array implementation, serving as the DUT for its matrix multiplication error detection research\cite{libano2023efficient}.
Libano's implementation incorporates INT8 packing tricks and the DSP DDR technique, techniques that are absent in tinyTPU, thus resulting in a superior design with enhanced performance. Fig.\ref{fig:tpuv1}A shows a TPUv1 like $4\times 4$ systolic matrix engine.

\subsection{Drawbacks in Existing Implementations}

\begin{figure}
    \centering
    \includegraphics[width=1.0\linewidth]{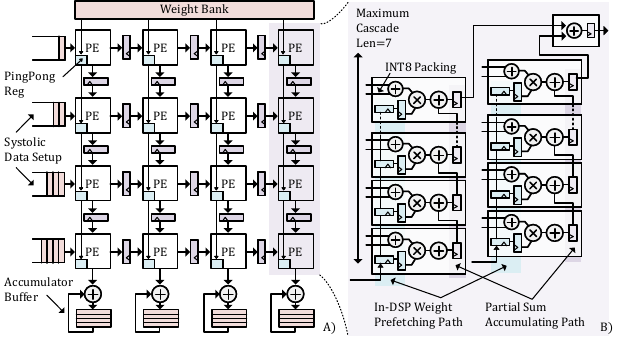}
    \caption{A) Google TPUv1-like systolic matrix engine. B) A single PE column of the proposed TPUv1-like systolic engine. \color{Cerulean}{The blue color in B indicates the weight prefetching path that is completely absorbed into the DSP48E2.}}
    \label{fig:tpuv1}
    \end{figure}

Despite the popularity of tinyTPU and the DSP-centric considerations in Libano's design, both implementations still fall short of achieving optimal performance.
tinyTPU does not employ the common INT8 packing techniques, resulting in half the computing density. Furthermore, activations at each row of the systolic array are broadcast to all columns of DSP48E2, instead of using pipelining registers. This approach leads to high fan-out and negatively impacts the frequency performance.
Libano's implementation fails to absorb the partial sums accumulating path into the DSP48E2, leading to excessive consumption of CLB resources.

In our implementation shown in Fig.\ref{fig:tpuv1}B, we integrate a comprehensive set of techniques, including INT8 packing and in-DSP partial sums cascading. Furthermore, we have identified a critical yet often neglected aspect, the weight loading datapath, that could potentially become the performance bottleneck for the WS systolic array. Recognizing this, we introduce an innovative in-DSP operand prefetching technique designed to optimize the weight loading datapath. 

\subsection{Enhancement: In-DSP Operand Prefetching}

In the WS systolic array, weights need to be preloaded into the array. To hide the preloading latency, each PE of the WS systolic array must instantiate two sets of registers. This setup enables the ping-pong update for loading the next set of weights.
In ASIC design, there's little room for optimizing register consumption. However, in FPGA design, it is possible to utilize existing resources for the ping-pong weight loading path rather than instantiate extra CLB flip-flops.

\begin{figure}
    \centering
    \includegraphics[width=1.0\linewidth]{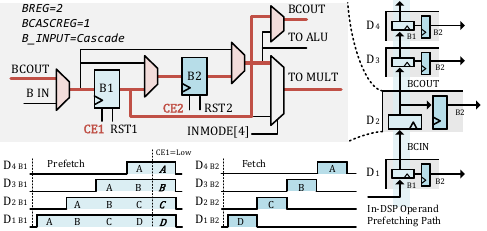}
    \caption{The proposed in-DSP operand prefetching technique. The cascaded $B_1$ registers form the shared weight prefetching path. When data stored in the $B_2$ registers expires, the data in the $B_1$ registers shifts into $B_2$.}
    \label{fig:indspfetch}
    \end{figure}

The DSP48E2 has two flexible input pipelines for input ports A and B, each comprising two registers with individual clock enables. These input pipelines can accept individual input from the general routing resources or share the same cascading path.
In this section, we demonstrate how to perform in-DSP operand prefetching by absorbing the weight ping-pong registers into the DSP48E2 pipelines, shown in Fig.\ref{fig:indspfetch}.

In an FPGA-based WS systolic array, the accumulating datapath can be absorbed into the vertical DSP48E2 output cascading path.
The unused input cascaded path can be utilized for weight prefetching. Assuming the input pipeline for B is used for weight prefetching, Fig.\ref{fig:indspfetch} illustrates the static configuration of the pipeline.
In this setup, $B_1$ registers serve as the operand prefetching path, while $B_2$ registers remain static across multiple computation rounds. New operands stream into the $B_1$ register chain until they reach the topmost DSP48E2. When operands cached in the $B_2$ registers expire, the clock enable signals trigger $B_2$ to accept new operands from the $B_1$ chain. The waveform of the clock enable signals for $B_1$ and $B_2$ registers is presented Fig.\ref{fig:indspfetch}.
The entire operand prefetching process, along with the partial sums accumulating path, occurs entirely within the DSP, flowing vertically down the DSP48E2 column. This results in significant savings in CLB flip-flops.

\subsection{Experiments}

We compare our implementation with tinyTPU and Libano's implementation under Vivado's out-of-context mode. This mode allows us to independently assess the performance of the systolic matrix engine without the interference of other components.
Table.\ref{tbl:tpuutil} shows that our proposed implementation with the in-DSP operand prefetching technique, significantly reduces the usage of LUTs and FFs compared to Libano's implementation, while also achieving a considerable improvement in clock frequency compared to the tinyTPU.

\begin{table}[]

\caption{Resource Util. Comparison of INT8 $14\times 14$ TPUv1 on XCZU3EG}
\label{tbl:tpuutil}
\begin{adjustbox}{max width=1.0\linewidth}
\begin{threeparttable}[b]
\begin{tabular}{l|l|l|l|l|l|l|l}
\hline
& LUT   & FF      & CARRY & DSP & Freq. & WNS   & Pow. \\ \hline\hline
tinyTPU\tnote{1} & 120      &  129       &   0     & 196    &  400     & 0.076      & 0.25      \\ \hline
Libano\tnote{2}  & 23080 & 60422   & 2734   & 196 & 666   & 0.044 & 4.87 \\ \hline
CLB-Fetch\tnote{3}    & 168   & 6195    & 0      & 210 & 666   & 0.083 & 0.94  \\ \hline                    
DSP-Fetch\tnote{3}    & 167   & 4516    & 0      & 210 & 666   & 0.052 & 0.93  \\ \hline
\end{tabular}
\begin{tablenotes}
\item[1] The MAC/DSP count of tinyTPU is actually half that of Libano's and ours, since tinyTPU does not incorporate the INT8 packing technique. The minimal usage of LUTs and FFs in tinyTPU is due to its data broadcasting instead of staging in the systolic array, which results in poorer frequency performance.
\item[2] The extensive usage of LUTs and FFs in Libano's implementation is due to the use of DDR Mux for all PEs and the employment of a CLB-based accumulating chain.
\item[3] CLB-Fetch follows the same implementation strategy as DSP-Fetch except that the weight fetching path is not integrated in the DSP.
\end{tablenotes}
\end{threeparttable}
\end{adjustbox}
\centering
\end{table}

\section{Enhancing Systolic Engine of Xlinx DPU}

\begin{figure*}
    \centering
    \includegraphics[width=1.0\linewidth]{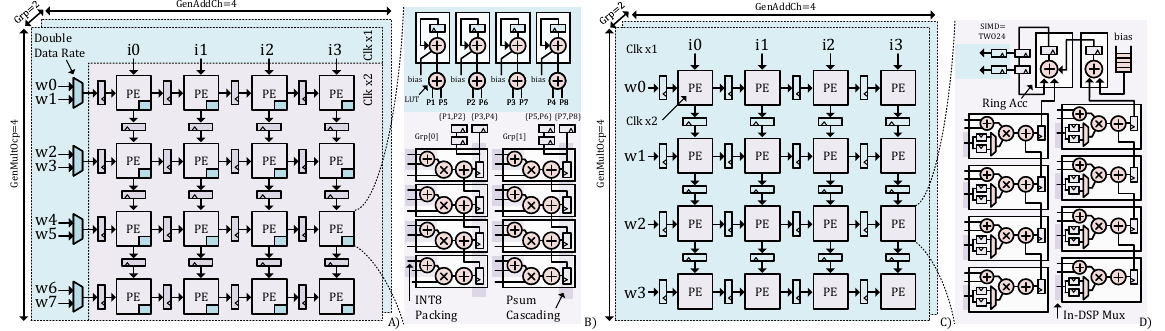}
    \caption{A) B1024 systolic engine of the DPUCZDX8G . B) The PE of DPUCZDX8G's systolic engine. C) Proposed enhanced systolic matrix engine. D) The PE of the proposed enhanced systolic engine. \color{Cerulean}{The blue color indicates the $Clk_{\times 1}$ clock domain}, \color{Mulberry}{while the purple color indicates the $Clk_{\times 2}$ clock domain.}}
    \label{fig:xilinxdpu}
    \end{figure*}       

Xilinx developed the Vitis AI DPU as a machine learning solution for FPGA\cite{kathail2020xilinx}.
DPUCZDX8G is the DPU IP core of the Vitis AI for the Zynq UltraScale series FPGA platforms\cite{xilinx2021dpu}.
We begin by providing a brief introduction to the systolic architecture of the DPUCZDX8G. Despite the encryption of the DPUCZDX8G IP core, the documentation for DPUCZDX8G contains ample information about its architecture. Furthermore, the utilization reports and the actual implementation layout obtained in Vivado IDE have verified the accuracy of the public documented description.

DPUCZDX8G employs an output stationary (OS) systolic engine with three levels of parallelism, involving input output channel and pixel parallelism. Fig.\ref{fig:xilinxdpu}A shows the B1024 configuration of DPUCZDX8G.
Each PE shown in Fig.\ref{fig:xilinxdpu}B comprises a group of DSP48E2 chains performing vector inner product computation. This PE design fully leverages the DSP48E2's dedicated cascaded path to reduce routing complexity.
Bundles of activations flow vertically into the systolic engine, bundles of weights flow horizontally into the systolic engine, and the products continually accumulate in the PE accumulator.

DPUCZDX8G also incorporates a DDR technique, allowing the DSP48E2s to operate at twice the clock rates of other logic components. CLB multiplexers switch between two data portions from a low-speed clock domain, delivering one portion at a time to the high-speed DSP48E2.
Dot products generated by the DSP chain are transferred back to the slow clock domain through a set of flip-flops that perform the serial-to-parallel conversion. These results are then accumulated by the low-speed accumulator.
This clock domain decoupling design maximizes the utilization of the DSP48E2's exceptional Fmax performance, eliminating any lag in the low-speed fabric.

\subsection{Drawbacks in DPUCZDX8G's Systolic Engine}

While Xilinx's officially-developed systolic architecture of the DPUCZDX8G is already considered near-optimal and has demonstrated superior performance in Xilinx FPGA, it still exhibits several drawbacks as listed below:

1) The employment of CLB multiplexers inevitably consumes general routing resources and fabric logic. Furthermore, as CLB multiplexers span two clock domains, they impose stress on timing constraints. This may lead to a degradation of the clock frequency in the high-speed clock domain.

2) The cost associated with the DDR technique results in the harsh doubled bandwidth requirements for the weight data.

3) Each fast DSP48E2 chain requires two slow DSP48E2 accumulators in DPUCZDX8G. The Fmax performance of the DSP48E2 accumulator is underutilized, and the required number of accumulators in DPUCZDX8G is costly.

4) Extra LUTs and CARRY8s are required for the grouped partial sums combining and INT8 correction process.

In this section, we unveil several untapped DSP48E2 optimization potentials aimed at addressing these drawbacks.
We successfully absorb the CLB multiplexers introduced by DDR techniques into the DSP48E2, eliminating the need for excessive LUT usage.
We shift the burden of the doubled bandwidth requirement from the weight data to the output results, given that the output results' bandwidth is much smaller in the OS dataflow.
We move the DSP48E2 accumulator from the slow clock domain to the fast clock domain and halve the number of required accumulators without affecting the throughput, thereby significantly reducing DSP48E2 consumption.
We tackle these issues through two new DSP48E2 techniques outlined below and shown in Fig.\ref{fig:xilinxdpu}C and Fig.\ref{fig:xilinxdpu}D..

\subsection{Enhancement: In-DSP Multiplexing}

In DPUCZDX8G, each row of the systolic matrix engine shares the same weights. Two portions of weights from the slow clock domain need to be multiplexed into the fast clock domain and streamed into the systolic engine, progressing through one flip-flop stage per processing element.
In our approach, the need for LUT multiplexing is eliminated, and the staging registers operate at the slow clock domain. This not only significantly alleviates timing closure pressure but also reduces power consumption.

\begin{figure}
    \centering
    \includegraphics[width=1.0\linewidth]{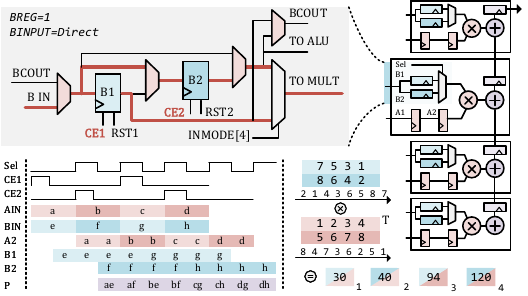}
    \caption{The proposed in-DSP multiplexing technique.}
    \label{fig:indspmux}
    \end{figure}

We make use of the two flexible input pipelines to perform in-DSP multiplexing. In the DSP48E2's input pipelines, the pathway from the input port to the multiplier can be configured statically through attribute settings or dynamically switched using a multiplexer.
For simplification in the illustration, we focus on the multiplication between input port A and input port B without any data packing process.
While data stream into the DSP48E2 from the slow clock domain, the DSP48E2 itself operates at a doubled data rate. We designate the slow clock as $Clk_{\times 1}$ and the fast clock as $Clk_{\times 2}$.
As shown in Fig.\ref{fig:indspmux}, input port A is set up with a simple 2-stage pipeline. Activations are streamed into the $A_1$ and $A_2$ registers. The clock enable pins for $A_1$ and $A_2$ are consistently high, and new data updates $A_1$ and $A_2$ every $Clk_{\times 1}$ cycle.
Conversely, input port B is configured with a ping-pong datapath. Weights are streamed into $B_1$ and $B_2$ registers in a ping-pong manner, controlled by the independent clock enable pins for $B_1$ and $B_2$. New data updates $B_1$ or $B_2$ every two $Clk_{\times 1}$ cycles.

The multiplexer in the input port B pipeline plays a crucial role: it switches between $B_1$ and $B_2$ registers at the speed of $Clk_{\times 2}$. This enables the multiplier to execute the cross product between activations $a_{t}, a_{t+1}$ and weights $w_{t}, w_{t+1}$ in adjacent $Clk_{\times 1}$ cycles. As a result, it yields $a_{t}w_{t}, a_{t}w_{t+1}, a_{t+1}w_{t}, a_{t+1}w_{t+1}$, four results every two $Clk_{\times 1}$ cycles, achieving DDR multiplication.
The streaming activations and weights can be pre-arranged in such interleaved manner in advance.
All the aforementioned signals are depicted in the waveform in Fig.\ref{fig:indspmux} to provide a clearer illustration.

INT8 packing technique can also be applied in combination with the proposed in-DSP multiplexing techniques. Since the A plus D pre-adder datapath, which is also a 2-stage pipeline, has no effect on the ping-pong datapath of the B input port.
Furthermore, cascading is applicable. Cascading $N$ DSP48E2 units with such settings, combined with INT8 packing, results in a processing element capable of performing matrix multiplication between $4\times N$ activations and $N\times 2$ weights every two $Clk_{\times 1}$ cycles.
Our proposed DSP48E2 chain achieves an equivalent computing density to the DSP48E2 chain in DPUCZDX8G, but eliminating the need for LUT multiplexing and halving the weight data bandwidth at the same time.

\subsection{Enhancement: Ring Accumulator}

In DPUCZDX8G, each DSP48E2 chain generates two pair of independent INT18 partial sums every $Clk_{\times 2}$ cycle. The partial sums are transferred back to $Clk_{\times 1}$ domain by serial-to-parallel conversion.
The official implementation, where products generated by DSP48E2 chains in the same group are combined using LUT adder-tree and each combined partial sum is processed by an SIMD=ONE48 DSP48E2 accumulator (which adds an INT26 bias using the DSP pre-adder and produce 29-bit final results), is not optimal.
Dealing with INT18 partial sums, INT26 bias, and INT29 final results leads to underutilization of the 48-bit DSP48E2 accumulator.
Our implementation reduce both the bias precision and the precision of the accumulator to INT24. This adjustment results in only a minor loss in precision but allows for an efficient alignment with the SIMD=TWO24 feature of the DSP48E2.

\begin{figure}
    \centering
    \includegraphics[width=0.8\linewidth]{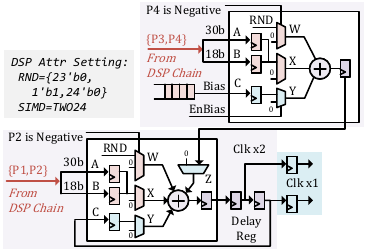}
    \caption{The proposed ring accumulator.}
    \label{fig:ringacc}
    \end{figure}

In our in-DSP multiplexing method, each DSP48E2 chain produces four pairs of INT18 partial sums every four $Clk_{\times 2}$ cycles.
Rather than transferring the partial sums back to the $Clk_{\times 1}$ domain and utilizing LUT adders for group combining, we design a ring accumulator composed of only two cascaded DSP48E2s that also operate at $Clk_{\times 2}$, handling the combining of partial sums, the addition of bias, and the accumulation process altogether, as depicted in Fig.\ref{fig:ringacc}.
The two groups, four pairs of partial sums sending to the ring accumulator are accumulated sequentially. The accumulator introduces a latency of two, and the outputs of the accumulator undergo a delay using two registers. The delayed outputs are connected back to the DSP48E2, aligning with the next iteration of accumulation. By adopting this approach, the number of DSP48E2 accumulators is effectively reduced by half.

It's noteworthy that the correction required by the INT8 packing does not demand extra LUTs or CARRY8 logic. We leverage the rounding constant at the $W$ multiplexer inside the DSP48E2 to handle the compensation. It is also worth noting that the delay registers in the ring accumulator loop can be repurposed for the serial-to-parallel conversion to transfer the accumulated results back to $Clk_{\times 1}$ domain, shown in Fig.\ref{fig:ringacc}.

In contrast to DPUCZDX8G's accumulator, our accumulator generates four pairs of sums instead of two. This difference is due to the burden of doubled weight bandwidth in DPUCZDX8G's implementation, now placed on the output. This is not expected to become a performance bottleneck as the output bandwidth, amortized over time, is generally small in OS dataflow, particularly considering that accumulation cycles are typically large in convolutional neural networks.

\subsection{Experiments}

\begin{figure}
    \centering
    \includegraphics[width=1.0\linewidth]{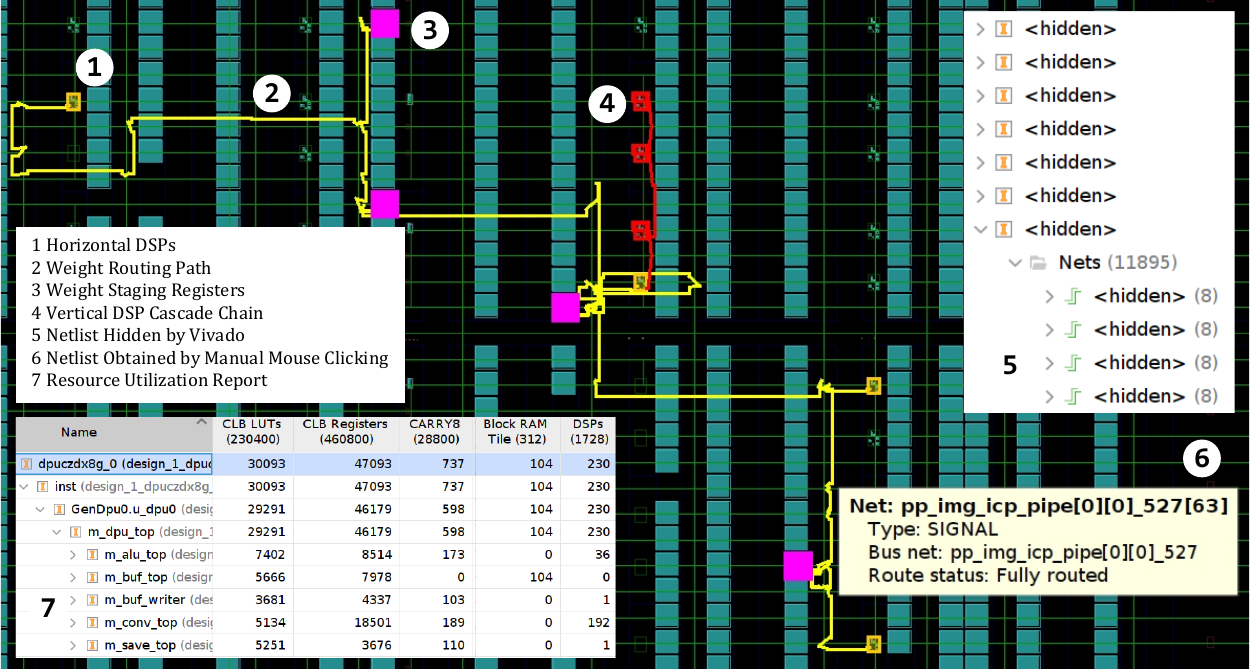}
    \caption{Analyzing DPUCZDX8G in Vivado post implementation GUI.}
    \label{fig:screnn}
    \end{figure}

DPUCZDX8G is a commercially encrypted IP, making it impossible to access its original module-level breakdown design directly for experimentation.
Although the module hierarchy of the encrypted DPUCZDX8G is explicitly hidden, it can still be obtained from the Vivado resource utilization report.
While the netlists of the encrypted DPUCZDX8G are also explicitly hidden, they remain implicitly visible in the Vivado implementation device view GUI when selecting the physical routing wire, shown in Fig.\ref{fig:screnn}. Additionally, we can still use the Vivado \textit{find} function to count cells or nets with the same naming prefix.
Through these methods and existing public documentation of the DPUCZDX8G, we were able to unravel the intricate design of the Vitis AI DPU. We obtain the resource utilization breakdown of the DPUCZDX8G systolic engine in B1024 configuration, shown in Table.\ref{tbl:dpuutil}.

To ensure a fair assessment solely on the systolic architecture, we recreate a one-to-one systolic matrix engine of the DPUCZDX8G B1024 to compare with our proposed implementation under Vivado's out-of-context mode.
Our replication closely adheres to the resource utilization breakdown of the critical systolic array component of B1024, removing other non-critical components to simplify the comparison and align with our proposed implementation, ensuring a fair comparison.
As shown in Table.\ref{tbl:dpuutil}, our proposed implementation achieves a substantial 85\% and 20\% reduction in total LUTs and FFs compared to the official B1024 replicate. We have also managed to halve the number of DSP48E2 accumulators. Moreover, under the same frequency, we have achieved a 20\% reduction in power and gained much more timing margin.

\begin{table}[]
    \centering
    \begin{threeparttable}[b]
    \caption{Resource Util. Breakdown Comparison of DPU B1024 Impl.}
    \label{tbl:dpuutil}
    \begin{tabular}{c|c|c|c|c|c}
    \hline
                    & Official & Ours  &               & Official & Ours \\ \hline\hline
    WgtWidth        & 512b     & 512b  & AddTreeLUT    & 1152     & \bf{0}    \\ \hline
    ImgWidth        & 512b     & \bf{256b}  & AddTreeFF     & 1216     & \bf{0}    \\ \hline
    PsumWidth       & 2304b    & 2304b & AddTreeCarry  & 192      & \bf{0}    \\ \hline
    PsumFF          & 3456     & 3456  & TotalLUT\tnote{1}      & 1280     & \bf{158}  \\ \hline
    WgtImgFF        & 3072     & 3072  & TotalFF\tnote{1}       & 7856     & \bf{6208} \\ \hline
    MultDSP         & 128      & 128   & Freq.\tnote{1}         & 666M     & 666M      \\ \hline
    AccDSP          & 64       & \bf{32}    &  WNS\tnote{1}           &   0.095       &  \bf{0.116}        \\ \hline
    MuxLUT          & 128      & \bf{0}     &  Power\tnote{1}         & 1.03W    &\bf{0.826W}  \\ \hline
    \end{tabular}
    \begin{tablenotes}
        \item[1] While the other resource breakdown metrics are from B1024, the total LUT and FF usage, WNS and power metircs are derived from our replicate, which retains only the critical components from the official designs to ensure alignment with our proposed impl. Experiments are conducted on xczu3eg.
      \end{tablenotes}
    \end{threeparttable}
    \end{table}

\section{Applicability on SNN Accelerator}

\begin{figure}
    \centering
    \includegraphics[width=1.0\linewidth]{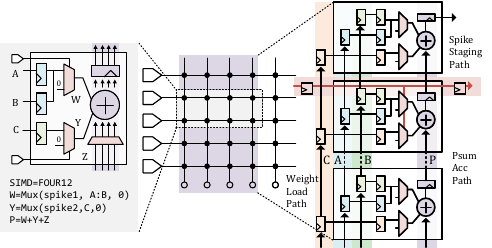}
    \caption{Applicability on SNN Accelerator. FireFly utilizes the multiplexers in DSP48E2 to perform spike-based computation. Half of the synaptic weights loading path can be absorbed in the input pipelines for A and B port.}
    \label{fig:snnsa}
    \end{figure}

In this section, we show that aforementioned techniques can seamlessly be applied to systolic-array-based SNN accelerators.
FireFly\cite{li2023fireflyv1} represents a state-of-the-art SNN accelerator, employing a typical WS systolic array design. It leverages the wide-bus multiplexers in DSP48E2 units for spike-based computation.
In FireFly's implementation, two sets of synaptic weights are presented on the A B port and the C port, respectively, with the weights on the A B port being concatenated within the DSP48E2, as illustrated in Fig.\ref{fig:snnsa}. Utilizing the SIMD=FOUR12 mode of the DSP48E2 allows a single DSP48E2 unit to function as a $2\times 4$ synaptic crossbar.

Our in-DSP operand prefetching technique can use both the A and B input pipelines along with their cascaded paths for synaptic weight prefetching, shown in Fig.\ref{fig:snnsa}.
While the use of CLB flip-flops is unavoidable for weights presented on the C port due to the absence of C cascaded paths in DSP48E2, the overall requirement for weight ping-pong CLB registers is still greatly reduced. 
As indicated in Table.\ref{tbl:fireflyutil}, the total flip-flops consumption is reduced by half compared to the original implementation, accompanied by a noticeable drop in power consumption.

\begin{table}[]
    \caption{Resource Util. Comparison of FireFly Impl. on XCZU3EG}
    \label{tbl:fireflyutil}
    \begin{threeparttable}[b]
    \begin{tabular}{l|l|l|l|l|l}
    \hline
                        & LUTs & FFs         & DSP48E2s & Frequency & Power   \\ \hline\hline
    FireFly\tnote{1}    & 60  & 4344            & 64  & 666M  & 0.160W  \\ \hline
    Ours                & 60  & \bf{2296}       & 64  & 666M  & 0.153W  \\ \hline
    \end{tabular}
    \begin{tablenotes}
        \item[1] The parallelism is set to $32\times 32$, with 4 DSP48E2 chains of length 16 connected horizontally, acting as a $32\times 32$ synaptic crossbar.
        \end{tablenotes}
    \end{threeparttable}
    \centering
    \end{table}

\section{Conclusion}

In this paper, we revisit the functionalities of the DSP48E2, exploring its optimization potential and uncovering several underutilized techniques that can enhance the performance of systolic matrix engines on UltraScale series FPGAs. We discuss the in-DSP operand prefetching technique, in-DSP multiplexing technique, and the ring accumulator technique, which can be broadly applied to both WS and OS systolic arrays, including those used for neuromorphic SNN computing. We believe that our contibutions offer insights for researchers and developers aiming to build DSP-optimized hardware.

\section{Acknowledgement}

This work is supported by the Chinese Academy of Sciences Foundation Frontier Scientific Research Program (ZDBS-LY- JSC013). 
This work is part of the software-hardware codesigns research of the Brain-inspired Cognitive Engine (BrainCog)\cite{braincogweb}\cite{Zeng2023}.
We would also like to express our gratitude to Niansong Zhang from Cornell University for providing valuable suggestions on the paper.

\bibliographystyle{IEEEtran}
\bibliography{myIEEE,reference}

\begin{thebibliography}{10}
\providecommand{\url}[1]{#1}
\csname url@samestyle\endcsname
\providecommand{\newblock}{\relax}
\providecommand{\bibinfo}[2]{#2}
\providecommand{\BIBentrySTDinterwordspacing}{\spaceskip=0pt\relax}
\providecommand{\BIBentryALTinterwordstretchfactor}{4}
\providecommand{\BIBentryALTinterwordspacing}{\spaceskip=\fontdimen2\font plus
\BIBentryALTinterwordstretchfactor\fontdimen3\font minus \fontdimen4\font\relax}
\providecommand{\BIBforeignlanguage}[2]{{%
\expandafter\ifx\csname l@#1\endcsname\relax
\typeout{** WARNING: IEEEtran.bst: No hyphenation pattern has been}%
\typeout{** loaded for the language `#1'. Using the pattern for}%
\typeout{** the default language instead.}%
\else
\language=\csname l@#1\endcsname
\fi
#2}}
\providecommand{\BIBdecl}{\relax}
\BIBdecl

\bibitem{xilinx2021ultrascale}
Xilinx Inc., ``Ultrascale architecture dsp slice user guide,'' 2021.

\bibitem{fu20178}
Yao Fu, Ephrem Wu, and Ashish Sirasao, ``8-bit dot-product acceleration,'' \emph{Xilinx Inc. San Jose, CA, USA}, p.~20, 2017.

\bibitem{han2020convolutional}
T~Han, Tianyu Zhang, Dong Li, Guangdong Liu, Lu~Tian, Dongliang Xie, and Yi~Shan Shan, ``Convolutional neural network with int4 optimization on xilinx devices,'' \emph{Xilinx White Paper, WP521}, 2020.

\bibitem{sommer2022dsp}
Jan Sommer, M~Akif {\"O}zkan, Oliver Keszocze, and J{\"u}rgen Teich, ``Dsp-packing: Squeezing low-precision arithmetic into fpga dsp blocks,'' in \emph{2022 32nd International Conference on Field-Programmable Logic and Applications (FPL)}.\hskip 1em plus 0.5em minus 0.4em\relax IEEE, 2022, pp. 160--166.

\bibitem{zhang2023uint}
Jingwei Zhang, Meng Zhang, Xinye Cao, and Guoqing Li, ``Uint-packing: Multiply your dnn accelerator performance via unsigned integer dsp packing,'' in \emph{2023 60th ACM/IEEE Design Automation Conference (DAC)}.\hskip 1em plus 0.5em minus 0.4em\relax IEEE, 2023, pp. 1--6.

\bibitem{li2023fireflyv1}
Jindong Li, Guobin Shen, Dongcheng Zhao, Qian Zhang, and Yi~Zeng, ``Firefly: A high-throughput hardware accelerator for spiking neural networks with efficient dsp and memory optimization,'' \emph{IEEE Transactions on Very Large Scale Integration (VLSI) Systems}, 2023.

\bibitem{jouppi2017datacenter}
Norman~P Jouppi, Cliff Young, Nishant Patil, David Patterson, Gaurav Agrawal, Raminder Bajwa, Sarah Bates, Suresh Bhatia, Nan Boden, Al~Borchers \emph{et~al.}, ``In-datacenter performance analysis of a tensor processing unit,'' in \emph{Proceedings of the 44th annual international symposium on computer architecture}, 2017, pp. 1--12.

\bibitem{xilinx2021dpu}
Xilinx Inc., ``Dpuczdx8g for zynq ultrascale+ mpsocs,'' 2021.

\bibitem{chen2021hardware}
Xiang Chen, Jindong Li, and Yong Zhao, ``Hardware resource and computational density efficient cnn accelerator design based on fpga,'' in \emph{2021 IEEE International Conference on Integrated Circuits, Technologies and Applications (ICTA)}.\hskip 1em plus 0.5em minus 0.4em\relax IEEE, 2021, pp. 204--205.

\bibitem{lee2018double}
Sugil Lee, Daewoo Kim, Dong Nguyen, and Jongeun Lee, ``Double mac on a dsp: Boosting the performance of convolutional neural networks on fpgas,'' \emph{IEEE Transactions on Computer-Aided Design of Integrated Circuits and Systems}, vol.~38, no.~5, pp. 888--897, 2018.

\bibitem{samajdar2019scaling}
Ananda Samajdar, Tushar Garg, Tushar Krishna, and Nachiket Kapre, ``Scaling the cascades: Interconnect-aware fpga implementation of machine learning problems,'' in \emph{2019 29th International Conference on Field Programmable Logic and Applications (FPL)}.\hskip 1em plus 0.5em minus 0.4em\relax IEEE, 2019, pp. 342--349.

\bibitem{sheet2017virtex}
Virtex UltraScale FPGAs~Data Sheet, ``Virtex ultrascale fpgas data sheet: Dc and ac switching characteristics ds893,'' \emph{San Jose, CA, USA}, 2017.

\bibitem{li2024fireflyv2}
Jindong Li, Guobin Shen, Dongcheng Zhao, Qian Zhang, and Yi~Zeng, ``Firefly v2: Advancing hardware support for high-performance spiking neural network with a spatiotemporal fpga accelerator,'' \emph{IEEE Transactions on Computer-Aided Design of Integrated Circuits and Systems}, 2024.

\bibitem{wu2017high}
Ephrem Wu, Xiaoqian Zhang, David Berman, and Inkeun Cho, ``A high-throughput reconfigurable processing array for neural networks,'' in \emph{2017 27th International Conference on Field Programmable Logic and Applications (FPL)}.\hskip 1em plus 0.5em minus 0.4em\relax IEEE, 2017, pp. 1--4.

\bibitem{tinytpugithub}
\BIBentryALTinterwordspacing
github. [Online]. Available: \url{https://github.com/jofrfu/tinyTPU}
\BIBentrySTDinterwordspacing

\bibitem{arora2021tensor}
Aman Arora, Samidh Mehta, Vaughn Betz, and Lizy~K John, ``Tensor slices to the rescue: Supercharging ml acceleration on fpgas,'' in \emph{The 2021 ACM/SIGDA International Symposium on Field-Programmable Gate Arrays}, 2021, pp. 23--33.

\bibitem{yang2022bp}
Jianchao Yang, Mei Wen, Junzhong Shen, Yasong Cao, Minjin Tang, Renyu Yang, Jiawei Fei, and Chunyuan Zhang, ``Bp-im2col: Implicit im2col supporting ai backpropagation on systolic arrays,'' in \emph{2022 IEEE 40th International Conference on Computer Design (ICCD)}.\hskip 1em plus 0.5em minus 0.4em\relax IEEE, 2022, pp. 415--418.

\bibitem{he2020sparse}
Xin He, Subhankar Pal, Aporva Amarnath, Siying Feng, Dong-Hyeon Park, Austin Rovinski, Haojie Ye, Yuhan Chen, Ronald Dreslinski, and Trevor Mudge, ``Sparse-tpu: Adapting systolic arrays for sparse matrices,'' in \emph{Proceedings of the 34th ACM international conference on supercomputing}, 2020, pp. 1--12.

\bibitem{libano2023efficient}
Fabiano Libano, Paolo Rech, and John Brunhaver, ``Efficient error detection for matrix multiplication with systolic arrays on fpgas,'' \emph{IEEE Transactions on Computers}, 2023.

\bibitem{kathail2020xilinx}
Vinod Kathail, ``Xilinx vitis unified software platform,'' in \emph{Proceedings of the 2020 ACM/SIGDA International Symposium on Field-Programmable Gate Arrays}, 2020, pp. 173--174.

\bibitem{braincogweb}
\BIBentryALTinterwordspacing
``Braincog: Brain-inspired cognitive intelligence engine.'' [Online]. Available: \url{http://www.brain-cog.network}
\BIBentrySTDinterwordspacing

\bibitem{Zeng2023}
\BIBentryALTinterwordspacing
Yi~Zeng, Dongcheng Zhao, Feifei Zhao, Guobin Shen, Yiting Dong, Enmeng Lu, Qian Zhang, Yinqian Sun, Qian Liang, Yuxuan Zhao, Zhuoya Zhao, Hongjian Fang, Yuwei Wang, Yang Li, Xin Liu, Chengcheng Du, Qingqun Kong, Zizhe Ruan, and Weida Bi, ``{BrainCog}: A spiking neural network based, brain-inspired cognitive intelligence engine for brain-inspired {AI} and brain simulation,'' \emph{Patterns}, p. 100789, Jul. 2023. [Online]. Available: \url{https://doi.org/10.1016/j.patter.2023.100789}
\BIBentrySTDinterwordspacing

\end{thebibliography}

\end{document}